\documentstyle[12pt,aasms4] {article}
\newcommand{\beq}{\begin{equation}}
\newcommand{\eeq}{\end{equation}}
\newcommand{\beqn}{\begin{eqnarray}}
\newcommand{\eeqn}{\end{eqnarray}}

\newcommand{\pa}{\partial}

\begin{document}
\title{\bf{On the Chaotic Orbits of 
Disc-Star-Planet Systems}}
\author{Ing-Guey Jiang $^{1}$ and Li-Chin Yeh$^{2}$}

\affil{
{$^{1}$Institute of Astronomy,}\\
{ National Central University, Chung-Li, Taiwan}\\
{$^{2}$Department of Mathematics,}\\
{ National Hsinchu Teachers College, Hsin-Chu, Taiwan} \\
}

\authoremail{jiang@astro.ncu.edu.tw}

\begin{abstract}
Following Tancredi, S\`anchez and Roig (2001)'s criteria
of chaos, 
two ways of setting initial velocities are used in the 
numerical surveys to explore the possible chaotic and regular orbits 
for the disc-star-planet systems. We find that
the chaotic boundary does not depend much on the disc mass for   
Type I initial condition, 
but can change a lot for different disc masses for 
Type II initial condition. A few sample orbits are further studied.
Both Poincar\'e surface of section 
and the Lyapounov Exponent
Indicator are calculated and 
they are consistent with each other.
We also find that the influence from the disc 
can change the locations of equilibrium points and the orbital
behaviors for both types of initial conditions. Because the chaotic orbits
are less likely to become the stable resonant orbits, we conclude
that the proto-stellar disc shall play important roles for the capture and 
depletion histories of resonant orbits of both Asteroid Belt and 
Kuiper Belt during the formation of Solar System.

\end{abstract}


\newpage
\section{Introduction}

In the recent years, more than 100 extra-solar planetary systems are 
discovered and most of them exhibit interesting features. 
Many results have been presented in the dynamical studies
trying to explain the observed orbital properties (for example, 
Laughlin \& Adams 1999, Rivera \& Lissauer 2000, Jiang \& Ip 2001, 
Ji et al. 2002). Among these previous works, 
the restricted three-body problem is often used as a model for these 
planetary systems.   

On the other hand, some of the discovered extra-solar planetary systems
are claimed to have discs of dust and these discs are regarded as the 
young analogues of the Kuiper Belt. For instance, Greaves et al. (1998) 
found a dust ring around a nearby star $e$ Eri 
and Jayawardhana et al. (2000) detected the dust in the 55 Cancri planetary 
system. Particularly, $\beta$ Pictoris planetary system has a warped
disc and the influence of a planet might explain this warp 
(Augereau et al. 2001). Jiang \& Yeh (2003) studied the effect of discs 
on the planetary orbits and concluded that the planets might prefer to stay
near the inner part rather than the outer part of the disc.  

Therefore, 
the modified restricted three-body problem proposed in 
Yeh \& Jiang (2003) is interesting and might have important astronomical
applications. In that model, the standard 
restricted three-body problem is generalized to the situation 
that the test particles are influenced by the additional 
gravitational forces from the disc.
The influence from the disc makes the structure of equal-potential surfaces
much more complicated and thus, the new equilibrium points
exist in addition to the usual Lagrangian points.
The conditions for the existence of these new equilibrium
points were studied both analytically and numerically in that paper. 

Having found the new equilibrium points, Jiang \& Yeh (2004a, 2004b)
studied the orbital behavior for this modified restricted three-body
problem. The sensitivities to the initial conditions for some sample orbits 
clearly depend on the disc mass.

In this paper, we focus on the case that the central binary 
is a Sun-Jupiter system for the modified restricted three-body problem.
Thus, it is a disc-star-planet system with the test particle moving around.
Although the current masses of Asteroid Belt and Kuiper Belt in the Solar 
System are small,
 it could be much more massive during the early stage. Thus, our 
study might  have important implications for the formation of 
the dynamical structure
near asteroid belts of any planetary systems.
Moreover, the chaotic behaviors of the test particles in this system might
be essential.
Indeed, the chaotic orbits do play roles for planetary dynamical properties. 
For example, Wisdom (1983) successfully showed that the 3:1 Kirkwood gap
of the asteroid belt might be related to the chaotic region of the system.
Moreover, Astakhov et al. (2003) used the model of three-body interaction
to explain the irregular satellites of the Jupiter. The possible chaotic
orbits were claimed to be the reason for some Jupiter' satellites
having irregular orbits.  

We present our models in Section 2. Two types of initial conditions will
be described in Section 3. The method to calculate the Lyapounov Exponent
Indicator is summarized in Section 4. The results will be in Section 5 and 6
and Section 7 concludes the paper.
 
\section{The Models}

We consider the motion of a test particle under the influence of 
the gravitational force of the star-planet system and the disc
with all of them in a two dimensional plane.

We assume that the two masses of the star-planet binary 
are $m_1$ and $m_2$ and choose
the unit of mass to make $G(m_1+m_2)=1$. If we define that 
$$\bar{\mu}=\frac{m_2}{m_1+m_2},$$
then 
the two masses are
$\mu_1=Gm_1=1-\bar{\mu}$ and $\mu_2=Gm_2=\bar{\mu}$.
The separation between the star and the planet 
is set to be unity. In this paper, we assume 
that $\mu_2=0.001 \ll \mu_1=0.999$. The location of the 
two masses are always at 
$(\mu_1, 0, 0)$ and  $(-\mu_2, 0, 0)$.
Because it is possible to have circular orbit for a planet moving under 
the influence of the central star and disc as shown in Jiang \& Yeh (2003),
our assumption that the star and planet are in circular motions is valid.
The only approximation is that the disc is centered on the star 
in Jiang \& Yeh (2003) but it is centered on the center of mass 
of the system in this paper.

The equation of motion of restricted three-body problem is 
(Murray \& Dermott 1999)
\beq\left\{
\begin{array}{ll}
&\frac{dx}{dt}=u \\
&\frac{dy}{dt}=v  \\
&\frac{du}{dt}=2v-\frac{\pa U^{\ast}}{\pa x}-\frac{\pa V}{\pa x}  \\
& \frac{dv}{dt}=-2u-\frac{\pa U^{\ast}}{\pa y}-\frac{\pa V}{\pa y},
\end{array} \right. \label{eq:3body1} 
\eeq
where

\beq
U^{\ast}=-\frac{1}{2}(x^2+y^2)-\frac{\mu_1}{r_1}-\frac{\mu_2}{r_2},
\label{eq:u_ast}
\eeq
\beq
r_1=\sqrt{(x+\mu_2)^2+y^2},\label{eq:r1}
\eeq
\beq 
 r_2=\sqrt{(x-\mu_1)^2+y^2}\label{eq:r2}
\eeq
and 
$V$ is the additional 
potential from the disc. We assume that $V$ is radially symmetric, 
so $V$ depends on the radial distance $r$, where $r={\sqrt {x^2+y^2}}$. Hence,
\beq\left\{
\begin{array} {ll}
&\frac{\pa V}{\pa x}= \frac{x}{r}\frac{\pa V}{\pa r}  \\
&\frac{\pa V}{\pa y}= \frac{y}{r}\frac{\pa V}{\pa r}, 
\end{array} \right.\label{eq:pav}
\eeq


We substitute Eq. (\ref{eq:u_ast}) and Eq. (\ref{eq:pav}) into 
Eq.(\ref{eq:3body1}) and have the following system:


\beq \left\{
\begin{array}{ll}
&  \frac{dx}{dt}=u  \\
& \frac{dy}{dt}=v  \\
& \frac{du}{dt}=2v +x-\frac{\mu_1(x+\mu_2)}{ r_1^3}-\frac{\mu_2(x-\mu_1)}
{r_2^3}-\frac{x}{r}\frac{\pa V}{\pa r}  \\
& \frac{d v}{dt}=-2u+y-\frac{y\mu_1}{r_1^3}-\frac{y\mu_2}{r_2^3}
-\frac{y}{r}\frac{\pa V}{\pa r},  
\end{array}  \right. \label{eq:3body2}
\eeq
where $r_1$ and $r_2$ are defined in Eq.(\ref{eq:r1})-(\ref{eq:r2}). 
We can consider any density 
profile of the disc by giving a formula of $V$ in Eq.
(\ref{eq:3body2}) as long as $V$ is radially symmetric.  In this paper, we
use Miyamoto-Nagai Potential for the disc. 

The analytic disc potential, from Miyamoto \& Nagai (1975),
is 
\beq
V(r,z)=-\frac{M_b}{\sqrt{r^2+(a+\sqrt{z^2+b^2})^2}},\label{eq:va1}
\eeq
where $M_b$ is the total mass of the disc and $r^2=x^2+y^2$. 
In this formula, $a$ and 
$b$ are two parameters which can change the disc profile slightly.
The parameter $a$ controls the flatness of the profile and can be called 
``flatness parameter''. The parameter $b$ controls the size of the core
of the density profile and can be called ``core parameter''.
When $a=b=0$, the potential equals to the one by a point mass. 

If we define $T\equiv a+b$, from Eq. (\ref{eq:va1}) we have
\beq
\frac{\pa V}{\pa r}(r,0)=\frac{M_b r}{(r^2+(a+b)^2)^{3/2}}=
\frac{M_b r}{(r^2+T^2)^{3/2}},\label{eq:va2}
\eeq
where we set $z=0$ since we only consider the orbits on the $x-y$ plane.
We substitute Eq. (\ref{eq:va2}) into Eq. (\ref{eq:3body2}) and have
the following system:
\beq \left\{
\begin{array}{ll}
&  \frac{dx}{dt}=u  \\
& \frac{dy}{dt}=v  \\
& \frac{du}{dt}=2v +x-\frac{\mu_1(x+\mu_2)}{r_1^3}-\frac{\mu_2(x-\mu_1)}
{r_2^3}-\frac{M_b x}{(r^2+T^2)^{3/2}}  \\
& \frac{d v}{dt}=-2u+y-\frac{y\mu_1}{r_1^3}-\frac{y\mu_2}{ r_2^3}
-\frac{M_b y}{(r^2+T^2)^{3/2}}.   
\end{array}  \right. \label{eq:ma3body2}
\eeq
We can see that neither the core parameter nor the flatness parameter
appears in the equations of motion. The dynamics of the system only 
depends on the summation of $a$ and $b$, i.e. $T$.

\section{Initial Conditions}

Since we consider a two dimensional model here, there are 
two variables, $x, y$ for space coordinates and another two variables,
$u, v$ for velocities to be set initially.  There are two types
of initial conditions:

For Type I,
we plan to set the initial 
position $(x_i,y_i)$ to be a grid point in a square, where $x_i \in [-2,2]$,
$y_i \in [-2,2]$ and with grid size 0.01. We will always set 
$u_i=v_i=0$ for the initial velocities of this type.
  
For Type II,
we plan to set the initial 
position $(x_i,y_i)$ to be a grid point in a square, where 
$x_i \in [-2.5,2.5]$,
$y_i \in [-2.5,2.5]$ and with grid size 0.01. Because,
for a test particle with radial coordinate $r_i$ in our system,
 the usual Keplerian 
circular velocity in the inertial frame is $v_c=\sqrt{1/r_i}$,
in our rotating frame, the circular velocity would be $v_c-r_i$.  
(Please note that 
both this test particle and the rotating frame rotate anti-clockwise and 
the angular speed of the rotating frame is set to be unity.)
We will set  
the initial velocities of this type as following:
Let $r_i={\sqrt {x_i^2+y_i^2}}$, $\theta_i=\tan^{-1}(y_i/x_i)$,
the initial velocity in the x-direction  
$u_i=-(\sqrt{1/r_i}-r_i)\sin\theta_i$;
the initial velocity in the y-direction  
$v_i=(\sqrt{1/r_i}-r_i)\cos\theta_i$.

The purpose to have Type I initial condition is that we wish
to study the orbital behavior near equilibrium points, where the velocity 
is always zero.  However, from 
Yeh \& Jiang (2003), we know that the spacial 
locations of equilibrium points will 
be different for different disc mass. By doing numerical surveys of this type
of initial conditions, we can always cover the region very close to the 
equilibrium points  for any disc mass.  

The purpose to have Type II initial condition is that we wish to make the 
initial orbits as the usual Keplerian orbits to model the early phase
of the planetary system.
These initial orbits are not circular due to the existence of the disc 
and the planet, though.

\section{Lyapounov Exponent Indicator}

For a complicated system
like ours, it is difficult to rigorously prove whether the orbits are chaotic 
or not. Nevertheless, we calculate the Lyapounov Exponent Indicator
to understand how sensitively the orbit depends  
on the initial conditions. 
According to Tancredi et al. (2001), using the two-particle method
to calculate  Lyapounov Exponent might lead to spurious results in certain
cases. Therefore,
we follow Wolf et al. (1985) to calculate Lyapounov Exponent Indicator
numerically.
To check if our calculation is correct, we have reproduced the results
of a given system in their paper.
We briefly summarize the procedure in  Wolf et al. (1985)
to calculate Lyapounov Exponent Indicator for a general dynamical system here.
We assume that a dynamical system is governed by below $n$ equations,
\beq
\frac{d y_i}{d t} = f_i(y_1, y_2,\cdots, y_n), \label{eq:non1}
\eeq
where $i=1, ..., n$ and the $n$ copies of 
linearized equations of this system are
\beq
\frac {d \overrightarrow{\bf w_{i}}}{d t} = {\bf J_f} 
\overrightarrow{\bf w_{i}}, 
\quad {\rm for} \,\, i=1,2,\cdots,n  \label{eq:lin1} 
\eeq
with the initial condition 
$\overrightarrow{\bf w_i}(0)=\overrightarrow {e_i}$, 
where ${\overrightarrow{\bf w_i}}=(w_{(1+n(i-1))},w_{(2+n(i-1))},\cdots,
w_{(n+n(i-1))})$, $\overrightarrow e_i$ is a unit vector in which
the $i$-th element equals to 1 and the rest are 0.
${\bf J_f}$ is a Jacobian matrix of $f_i$ and this Jacobian matrix depends
on $y_i$ for $i=1,\cdots,n$.

To calculate Lyapounov Exponent Indicator, we have to numerically solve System 
(\ref{eq:non1}) with chosen initial condition and System (\ref{eq:lin1})
simultaneously. One has to choose a sampling interval $\Delta t$ for rescaling.
(For this paper, 
we choose this sampling interval $\Delta t=1.0$.)
Every time when  $t/\Delta t$ is a positive integer, 
we record the length of vector  $\overrightarrow{\bf w_1(t)}$ and also
use Gram-Schmidt orthonormalization procedure on  
$\overrightarrow{\bf w_i(t)}$ to get $\overrightarrow{\bf w'_i(t)}$.
The  length of the vector $\overrightarrow{\bf w_1(t)}$ is proportional to 
$2^{\lambda_1 t}$, where $\lambda_1$ is the maximum  Lyapounov Exponent
at this stage and thus Lyapounov Exponent Indicator $\chi(t)$ is 
$\ln \|\overrightarrow{\bf w_1(t)}\|/(t\ln_2)$ where $\|\,\|$ is the length
of a given vector. We then use new set of vectors
$\overrightarrow{\bf w'_i(t)}$ as the new initial conditions for the
linearized  System (\ref{eq:lin1}) and continue to integrate both 
Systems (\ref{eq:non1}) and  (\ref{eq:lin1}) until next sampling time.  
The above process will be repeated until $t$ approaches a large number
since the maximum  Lyapounov Exponent 
$\gamma = {\rm lim}_{t\rightarrow\infty} \chi(t)$.

Then, the system is said to be 
chaotic if $\gamma >0 $ or otherwise regular. However, it is not possible 
to take $t \rightarrow \infty$ in practice. Thus, the evolution is followed 
up to some time and we usually plot $\ln \chi(t) $ as function of $\ln t$. 
According to Tancredi et al. (2001),
if the curve on this plot shows a negative  constant slope, the system is
regular, otherwise it is chaotic.  
We thus use the slope of the last part of the curve in 
$\ln (t)-\ln \chi(t)$ plane as the condition to be chaotic or regular.
(We ignore those points with negative $\chi(t)$.) 
That is, if 
\beq
[\ln \chi(t_{e2})- \ln \chi(t_{e1})]/[\ln(t_{e2}) - \ln(t_{e1})]< -0.5
\label{eq:condition}
\eeq 
for an orbit, then it is
regular, otherwise, it is chaotic, where  
$\ln \chi(t_{e1})$ ($\ln \chi(t_{e2})$) 
is the upper envelope points of the curve and 
the corresponding $\ln(t_{e1})$ ($\ln(t_{e2})$) is chosen to be
the one which is
 less than and closest to 5 (6.9). In the case that there is no such 
upper envelope point i.e. the curve is monotonic around there, 
we would then use $t_1$ ($t_2$) to 
replace $t_{e1}$ ($t_{e2}$),  where
$\ln(t_1)=5$ ($\ln(t_2)=6.9$), in Eq.(\ref{eq:condition}).





Moreover, if the test particle has been very far away from the central star,
i.e. ${\sqrt {x^2+y^2}}> 100$ at any time of the calculations,  
its orbit is called the ejected orbit.

\section{The Results for Type I Initial Condition}
Following the procedure in last section, we numerically solve 
System (\ref{eq:ma3body2}). 
Figure 1 is the result for the numerical survey of Type I initial condition.
We use different color to mark the initial locations of the test particles
which have particular orbital behavior as chaotic,  
regular-non-ejected or regular-ejected orbits.
The green region is the one with chaotic orbits, 
the red region is the one with regular-non-ejected orbits,
the blue region is  the one with regular-ejected orbits.
In general, every orbital calculation we present here is accurate to the 
level that the variation of the Jacobi's integral of the system
is less than three per cent of the initial value.
Because the orbital integrations for the test particles initially very close
to the central star and the planet are much more difficult and 
the calculations cannot satisfy the accuracy requirement, 
there are
small blank regions near the central star and the planet. 
The little white dots in the outer green belt of Figure 1(a) 
are just due to that the size
of the green points are not big enough to cover them under the resolution 
of our numerical survey.  There are also some
horizontal lines in each panel of Figure 1 which are due to the physical
size of the points. These are nothing to do with our results.
 
Please note that there are not any chaotic orbits which 
belong to ejected orbits  and all ejected orbits are regular.
It is natural that ejected orbits are regular since these orbits
become hyperbolic during ejections and the particles 
are far from the central object
with no further interaction with the system.

It is interesting that those test particles 
initially close to the central star would not be ejected
but those initially far away from the central star 
($r > 1.5$) would always get ejected.
This is probably because the test particles initially far away 
from the central star have larger  potential energy.
If it is located around $r=1$ initially, it could be chaotic or regular.
There is a belt of chaotic region 
around $r=1$ and another small belt around $r=0.1$. 
The belt of chaotic region around $r=1$ is probably due to the 
presence of the planet. Another belt of chaotic region around $r=0.1$
could be due to the multiple close interactions with the central star. 
Because both the potential and kinetic energy are very small 
for the test particles 
with small initial radial coordinates, they would stay around the region close 
to the central star for a longer time and thus move on chaotic orbits. 
For the test particles initially locate at larger radial coordinates, say 
the region between these two belts,
the potential energy is larger and thus they would gain some kinetic 
energy while they fall into the region of central star and then get back to 
the original larger radial coordinates again. Their interactions with 
the central star is therefore much less frequent.
Their orbits are not as chaotic
as those test particles
with initial radial coordinates around $r=0.1$ and thus are found to 
be regular. 



We find that these results are not sensitive to the disc mass or 
disc models. Nevertheless, below examples show that the 
locations of equilibrium points and the orbital behaviors 
will change with disc models. 

Figure 2 is the interesting example orbits for Type I initial condition.
There are four orbits in Figure 2 and all have initial condition:
$x=0.5-\mu_2,  y={\sqrt 3}/2, u=v=0$ but their disc potentials are different. 
(The initial location is marked by the small box in each panel of Figure 1.)

Figure 2(a) is 
the $\ln (t)-\ln \chi(t)$ plots for these orbits with 4 different
disc:
(1) the dashed curve is  the result of $M_b=0$, (2) 
the dotted curve is the result 
of $M_b=0.1$ and $a=b=0.5$ in  Miyamoto-Nagai Potential, 
(3) the dotted-dashed curve is the result 
of $M_b=0.1$ and $a=2.5, b=0.5$ in  Miyamoto-Nagai Potential
and (4)  the solid curve is the result 
of $M_b=0.1$ and $a=0.5, b=0.1$ in  Miyamoto-Nagai Potential.

 Figure 2(b) shows that
the first one is a stable orbit (as proved in Murry \& Dermott 1999) 
and it is only a small point in this panel.
The dotted curve in Figure 2(b) is the orbit in $x-y$ plane for disc (2)  
and it is a tadpole orbit. Figure 2(c) is the orbit for disc (3) and it is also
a tadpole orbit. Figure 2(d) is the one for disc (4) and it is a chaotic orbit,
which is confirmed by the solid curve of  $\ln (t)-\ln \chi(t)$ plot
in Figure 2(a).


\section{The Results for Type II Initial Condition}

Similarly, Figure 3
is the result for the numerical survey of Type II initial condition.
We use different color to mark the initial locations of the test particles
which have particular orbital behavior as chaotic
or regular orbits.
The green region is the one with chaotic orbits, 
and the red region is the one with regular orbits.
There are not any ejected orbits here.
Figure 3(a) is the results when there is no disc,
Figure 3(b)-(d) are the results when the disc mass is 0.1 $M_\odot$,
0.3 $M_\odot$ and 0.5 $M_\odot$ individually. The disc potential parameter 
$T=1$ for all of them.
It is obvious that the chaotic region becomes larger when we increase
the disc mass.
We find that there is no chaotic region near the central star for this 
kind of initial conditions.

Using Figure 3(b) as the guiding map, we further explore the orbital details
when the test particles are placed at different locations initially while
the disc mass is $0.1 M_\odot$ and $a=b=0.5$, i.e. $T=1$, in 
Miyamoto-Nagai Potential. These initial locations are marked by the small 
boxes in Figure 3(b).

For example, Figure 4 is the results of two different 
orbits with initial $(x,y)=(1.2,0)$
and $(x,y)=(2,0)$. Figure 4(a) is the  
$\ln (t)-\ln \chi(t)$ plots and Figure 4(b) is the Poincar\'e
surface of section for these two orbits.  The trajectories on $x-y$ plane
are in Figure 4(c)-(d).
The dotted curve of Figure 4(a) is for the orbit with initial condition
$(x,y)=(1.2,0)$ and the solid curve is for the orbit with initial condition
 $(x,y)=(2,0)$. It is clear that the dotted curve indicates that the orbit
is chaotic and the solid curve shows that its corresponding orbit is 
regular. The triangles in Figure 4(b) is the 
surface of section for the orbit with initial condition
 $(x,y)=(1.2,0)$ and these triangles spread over a large area of phase space.
The small ellipse is the surface of section 
for the orbit with initial condition $(x,y)=(2,0)$.
These results of the  Poincar\'e surface of section support the 
same conclusion as the one gained from $\ln (t)-\ln \chi(t)$ plots.
Finally, we plot the orbit with initial $(x,y)=(1.2,0)$ on $x-y$ plane
as Figure 4(c) and the orbit with initial $(x,y)=(2,0)$ on $x-y$ plane
as Figure 4(d). Indeed, the orbit in Figure 4(c) looks more randomly 
distributed than the orbit is Figure 4(d). 

When we check the guiding map, Figure 3(b), we find that 
$(x,y)=(2,0)$ lies in the red (regular) region  and $(x,y)=(1.2,0)$ 
is in the green (chaotic) region.  Therefore, our results in Figure 4
is completely consistent with the results of Figure 3(b).


Figure 5 is the result of two different orbits with initial 
$(x,y)=(0.8,0.55)$ and $(x,y)=(0.8,0.7)$. Figure 5(a) is the  
$\ln (t)-\ln \chi(t)$ plot and Figure 5(b) is the Poincar\'e
surface of section for these two orbits.  The trajectories on $x-y$ plane
are in Figure 5(c)-(d).
The solid curve of Figure 5(a) is for the orbit with initial condition
$(x,y)=(0.8,0.55)$ and the dotted curve is for the orbit with initial condition
 $(x,y)=(0.8,0.7)$. It is clear that the solid curve indicates that the orbit
is chaotic and the dotted curve shows that its corresponding orbit is 
regular. The triangles in Figure 5(b) is the 
surface of section for the orbit with initial condition
 $(x,y)=(0.8,0.55)$ 
and these triangles spread over a large area of phase space.
There is no any point on this surface of section for
the orbit with initial condition
 $(x,y)=(0.8,0.7)$ since this orbit does not go across $x-axis$, i.e. $y=0$.
These results of the  Poincar\'e surface of section  support the 
same conclusion as the one gained from $\ln (t)-\ln \chi(t)$ plots.
Finally, we plot the orbit with initial $(x,y)=(0.8,0.55)$ on $x-y$ plane
as Figure 5(c) and the orbit with initial $(x,y)=(0.8,0.7)$ on $x-y$ plane
as Figure 5(d). Indeed, the orbit in Figure 5(c) looks more randomly 
distributed than the orbit in Figure 5(d), which is actually a tadpole orbit. 

When we check the guiding map, Figure 3(b), we find that 
$(x,y)=(0.8,0.7)$ lies in the small red (regular) region  
as a gap of the green region and $(x,y)=(0.8,0.55)$ 
is in the green (chaotic) region.  Therefore, our results in Figure 5
are completely consistent with the results in Figure 3(b).


Moreover, Figure 6 is the results of two different 
orbits with initial $(x,y)=(-0.15,1.2)$
and $(x,y)=(-0.15,1.5)$. Figure 6(a) is the  
$\ln (t)-\ln \chi(t)$ plot and Figure 6(b) is the Poincar\'e
surface of section for these two orbits.  The trajectories on $x-y$ plane
are in Figure 6(c)-(d).
The solid curve of Figure 6(a) is for the orbit with initial condition
$(x,y)=(-0.15,1.2)$ 
and the dotted curve is for the orbit with initial condition
$(x,y)=(-0.15,1.5)$. It is clear that the solid curve indicates that the 
corresponding orbit
is chaotic.
From most part of this dotted curve
in $\ln (t)-\ln \chi(t)$ plot, it is likely to be a regular orbit.
The triangles in Figure 6(b) is the 
surface of section for the orbit with initial condition
 $(x,y)=(-0.15,1.2)$ 
and these triangles spread over a large area of phase space.
The two small elongated ellipses are the surface of section 
for the orbit with initial condition $(x,y)=(-0.15,1.5)$.
These results of the  Poincar\'e surface of section support the 
same conclusion as the one gained from $\ln (t)-\ln \chi(t)$ plots.
Finally, we plot the orbit with initial $(x,y)=(-0.15,1.2)$ on $x-y$ plane
as Figure 6(c) and the orbit with initial $(x,y)=(-0.15,1.5)$ on $x-y$ plane
as Figure 6(d). Indeed, the orbit in Figure 6(c) looks more randomly 
distributed than the orbit in Figure 6(d). 

When we check the guiding map, Figure 3(b), we find that 
$(x,y)=(-0.15,1.2)$ lies in the green (chaotic) region and 
$(x,y)=(-0.15,1.5)$ is in the red (regular) region 
Therefore, our results in Figure 6
are completely consistent with the results in Figure 3(b).


\section{Concluding Remarks and Implications}

We have studied the chaotic orbits for the disc-star-planet systems
through the calculations of Lyapounov Exponent 
and also Tancredi et al. (2001)'s criteria of chaos.
Two types of initial conditions are used in the 
numerical surveys to explore the possible chaotic and regular orbits 
and thus determine the boundaries. We find that
the chaotic boundary does not depend on the disc mass much for   
Type I initial condition, 
but can depend on the disc mass for 
Type II initial condition.
That is, discs with different masses might change the sizes and locations 
of chaotic region. 
Some sample orbits are further studied
and we find that the plots of Poincar\'e
surface of section are consistent with the results of Lyapounov Exponent
Indicator.

On the other hand, we notice that the influence of the disc 
can change the locations of equilibrium points and also the orbital
behaviors for both types of initial conditions. 
This point is particularly interesting. Without the influence from the 
disc, Wisdom (1983) showed that one of the Kirkwood gaps
might be explained by the chaotic boundary. This is probably 
due to that the chaotic orbits are less likely to become resonant orbits.
Indeed, we do find chaotic orbits in our disc-star-planet system 
and the disc, which can change orbital behaviors etc., shall play essential
roles during the formation of Kirkwood gaps. 
For example, 
the disc might make some test particles to transfer from chaotic orbits
to regular orbits
or from regular orbits to chaotic orbits
during the evolution.

The discs are often observed with various masses in both 
the young stellar systems and extra-solar 
planetary systems.
It is believed that the proto-stellar discs are formed during the early stage
of star formation and the planets might form on the discs through 
core-accretion or disc instability mechanism. The planetary 
formation time scale is not clear but the gaseous disc will be somehow 
depleted and 
the dust particles will gradually form. 
The above picture is plausible but unfortunately the time scales are 
completely unknown. For a system with a star, planets, discs and asteroids, 
it is not clear what is the order of different component's 
formation and what the duration of each component's formation process
would be.

Nevertheless, one thing we can confirm is that the discs are there and
can have different mass at different stages. From our results, it is clear
that the dynamical properties of orbits, i.e chaotic or regular,  
can be determined quickly 
for different disc masses or models. Because those particles moving
on chaotic orbits might have smaller probability to become the resonant objects
in the system, the Kirkwood gaps, which are associated with 
the location of mean motion resonant, in the Solar System might be formed 
by the dynamical effect we have studied here.
The detail processes would depend on the evolutionary 
time scale of the star, planet and also the disc.
To conclude, it is important that the effect of discs on the chaotic orbits
might influence the formation of the structures of planetary systems such as
the Kirkwood gaps in the Solar System.










\section*{Acknowledgment}
We are grateful to the referee,  Gonzalo Tancredi, for the excellent 
suggestions.
This work is supported in part 
by National Science Council, Taiwan, under 
Ing-Guey Jiang's Grants: NSC 92-2112-M-008-043 and also under Li-Chin Yeh's 
Grants: NSC 92-2115-M-134-001.
The numerical surveys are done by the PC Cluster, IanCu,
located in Institute of Astronomy, National Central University.


\end{document}